\documentstyle[12pt,epsfig,here]{article}
\begin{document}
\begin{sloppypar}
\begin{center}
{\bf SEARCH FOR MAGNETIC MONOPOLES WITH DEEP}
{\bf UNDERWATER CHERENKOV DETECTORS AT LAKE BAIKAL}

\vspace{4mm}
{\footnotesize
L.B.Bezrukov, B.A.Borisovets, I.A.Danilchenko, J.-A.M.Djilkibaev,
G.V.Domogatsky, A.A.Doroshenko, A.A.Garus, A.M.Klabukov, S.I.Klimushin,
B.K.Lubsandorzhiev, A.I.Panfilov, D.P.Petukhov, P.G.Pokhil, I.A.Sokalski \\
{\em Institute for Nuclear Research,  Russian Academy of Sciences (Moscow, 
Russia) \\}
\smallskip
N.M.Budnev, A.G.Chensky, V.I.Dobrynin, O.N.Gaponenko, O.A.Gress,
T.A.Gress, A.P.Koshechkin, R.R.Mirgazov, A.V.Moroz, S.A.Nikiforov,
Yu.V.Parfenov, A.A.Pavlov, K.A.Pocheikin, P.A.Pokolev, V.Yu.Rubzov,
S.I.Sinegovsky, B.A.Tarashansky \\
{\em Irkutsk State University (Irkutsk, Russia) \\}
\smallskip
S.B.Ignat$'$ev, L.A.Kuzmichov, N.I.Moseiko, E.A.Osipova \\
{\em Moscow State University  (Moscow, Russia) \\}
\smallskip
S.V.Fialkovsky, V.F.Kulepov, M.B.Milenin \\
{\em Nizhni Novgorod State Technical University  (Nizhni Novgorod, Russia) \\}
\smallskip
                                  M.I.Rozanov \\
{\em   St.Petersburg State Marine Technical University (St.Petersburg, 
Russia) \\}
\smallskip
                                  A.I.Klimov \\
{\em                     Kurchatov Institute (Moscow, Russia) \\}
\smallskip
                               I.A.Belolaptikov \\
{\em             Joint Institute for Nuclear Research  (Dubna, Russia) \\}
\smallskip
H.Heukenkamp$^*$, A.Karle, T.Mikolajski, Ch.Spiering,
               O.Streicher, T.Thon, Ch.Wiebusch, R.Wischnewski \\
{\em           DESY Institute for High Energy Physics (Zeuthen, 
Germany),  $^*$now at University of Wisconsin}
\vspace{3mm}
}
{\small
{\sf presented by I.Sokalski}

{\sf to the 2nd Workshop on 
"Dark Side of the Universe: Experimental efforts and theoretical framework",
Rome, Italy, November 13-14 1995}
}
\end{center}
\end{sloppypar}
\vspace{-4mm}
\begin{abstract}
{\footnotesize
The deep underwater Cherenkov neutrino telescope NT-200 is
currently under
construction at Lake Baikal. The "subdetectors" NT-36 (1993-95) and NT-72
(1995-96) have been operating successfully over 3 years. Various
techniques have been developed to search for magnetic monopoles with these
arrays. Here we describe a method used to detect
superheavy slowly moving ($\beta \approx 10^{-5} - 10^{-3}$)
monopoles
catalyzing baryon decay. We present results obtained  from  the
preliminary analysis of the data taken with NT-36 detector in 1993.
Furthermore, possibilities to observe faster ($\beta \approx 0.2  -  1$)
monopoles via other effects are
discussed.}
\end{abstract}
\vspace{1mm}
{\bf 1. The Baikal Deep Underwater Detectors}

\vspace{3mm}
The Baikal Neutrino Telescope$^1$
is being deployed
in Lake Baikal, 3.6 km from shore at the depth of 1.1  km.
In April 1993 we put into operation the stationary \mbox{3-string} 
detector  NT-36
carrying 36 PMTs. It took data  till  March  1995.  Since  April  1995  the
5-string detector NT-72 operates. These arrays are steps towards the
NT-200 detector which will consist of 192 optical modules (Fig.1  in $^2$).

The optical modules are grouped in pairs along the strings directed
alternatively upward and downward. The distance between pairs looking  face
to face is 7.5 m, while pairs arranged back to back are 5 m apart. The pulses
from two PMTs of a pair after
0.3 {\it p.e.} discrimination are fed to a coincidence circuit
with 15 ns time window. A PMT pair defines a {\it channel} with
its output denoted as
{\it local triggers} (or simply {\it hit}). The coincidence allows
to reduce the 1 {\it p.e.}
counting rate due to dark current noise and water luminescence by  more
then 2 orders, down to typically \mbox{50-500 Hz} 
depending on the season.

A {\it muon-trigger} is formed  by  the  requirement  of  $\geq3$ hit
channels within \mbox{500 ns.}  For  such  events,
amplitude and time of all fired channels  are  digitized  and
sent to shore. The event record includes all
hits within a time window of -1.0 $\mu$sec to +0.8 $\mu$sec
with respect to the  muon  trigger  signal. An independent
{\it monopole-trigger} system selects
candidate events
for slowly moving bright objects (see Sect. 2).
\vspace{5mm}
\begin{par}
\noindent
{\bf 2. Search for Monopoles Catalyzing Baryon Decay}
\vspace{2mm}
\end{par}

It was shown by Rubakov$^3$
and Callan$^4$  that
baryon number violation is possible in the presence of a GUT
magnetic monopole. For  reasonable  velocities  ($\beta \leq 10^{-3}$),
a catalysis cross section
$\sigma_c = 0.17 \cdot \sigma_o / \beta^{2}$
is predicted for monopole-proton interactions$^5$ with
$\sigma_o$ being of the order
of magnitude typical for strong interactions.
Following these predictions, the average
distances between two sequential  proton  decays  along the
monopole track in water can be as short as 10$^{-2}$ - 10$^{1}$ cm.

In order to search for GUT monopoles, a dedicated trigger system was
implemented in the electronics scheme. It is
based on the method which has been developed for
the first-stage Baikal setups GIRLYANDA$^6$ which operated
in 1984-89 (this  method  is  used  now  also  by the AMANDA
Collaboration$^7$  to search  for monopoles  deep
under ice). The method selects
events which are defined as a short-time
(0.1 - 1 msec) \mbox{1 {\it p.e.}} increase of the
counting  rate of individual
channels.
This pattern is expected from
sequential Cherenkov flashes produced by a  monopole along
it's track  via the proton  decay  products.  Due  to the large
$\sigma_c$ values,
for a short time interval the rate of detected
flashes can appreciably exceed the counting rate from PMT
dark current noise and  water luminescence,
even for monopoles
passing a channel at a distance of several tens of meters.

Our monopole system consists of several nearly
independent modules. They are synchronized
by a common
10 kHz clock.
One module reads the signals of 6
neighbouring channels placed
on the same string. During standard data taking runs, the
monopole trigger condition was defined as $\geq3$ hits  within
a time window of 500 $\mu$sec in  any  of  the  channels
(values for number of hits and time windows  duration  can
be set from shore).  Once the trigger  condition  is
fulfilled, the information about the number of hits
in each of the 6 channels which
occur within the given time window is
sent to shore. Time and amplitude information is
not recorded by the monopole system.

Since the method is based on the search for counting rate
excesses, it was important to perform a long-term {\it in situ}
check and to verify that the time behaviour
\begin{figure}[H]
\centering
\mbox{\epsfig{file=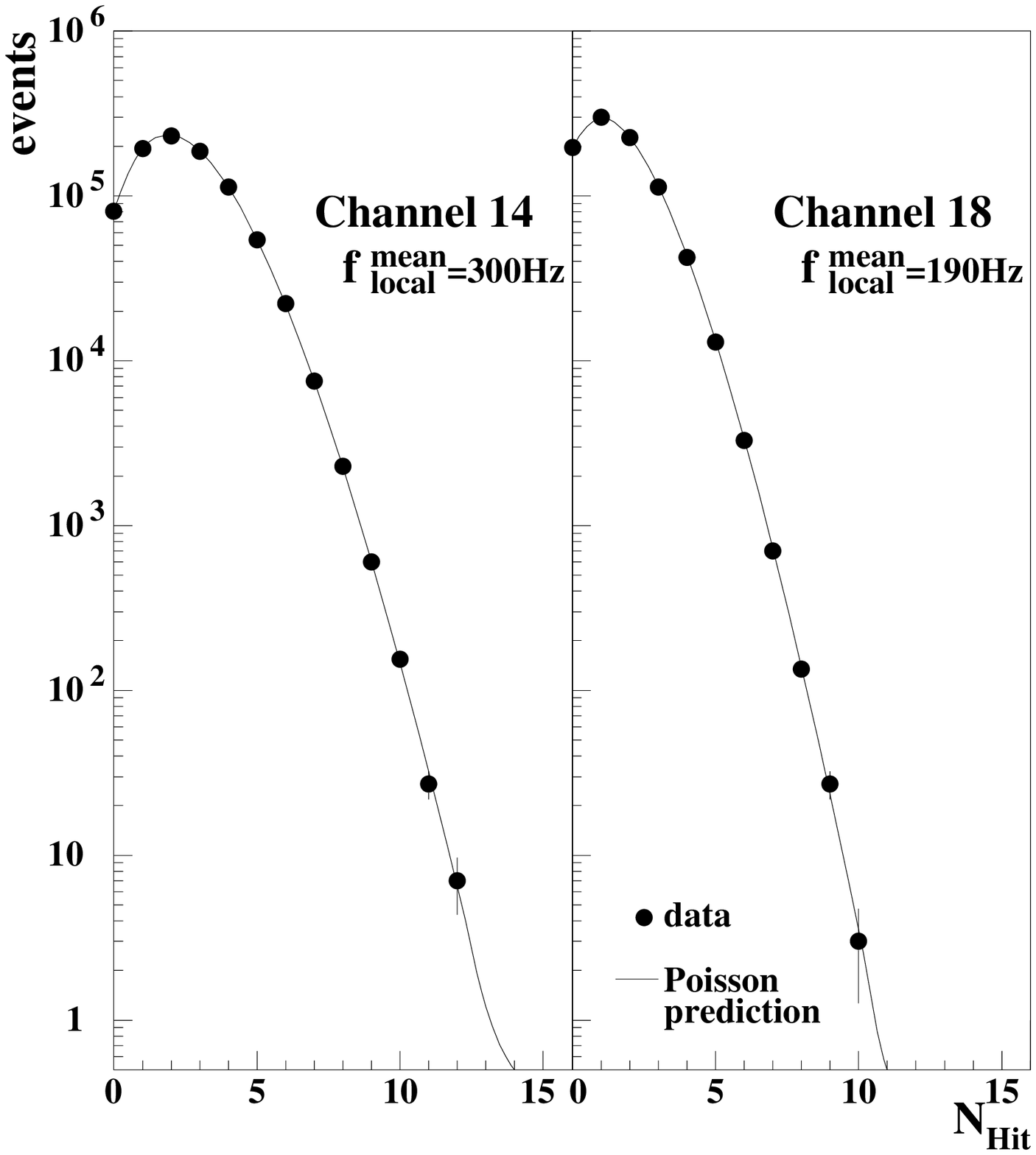,width=7.2cm,height=7.8cm}}
\mbox{\epsfig{file=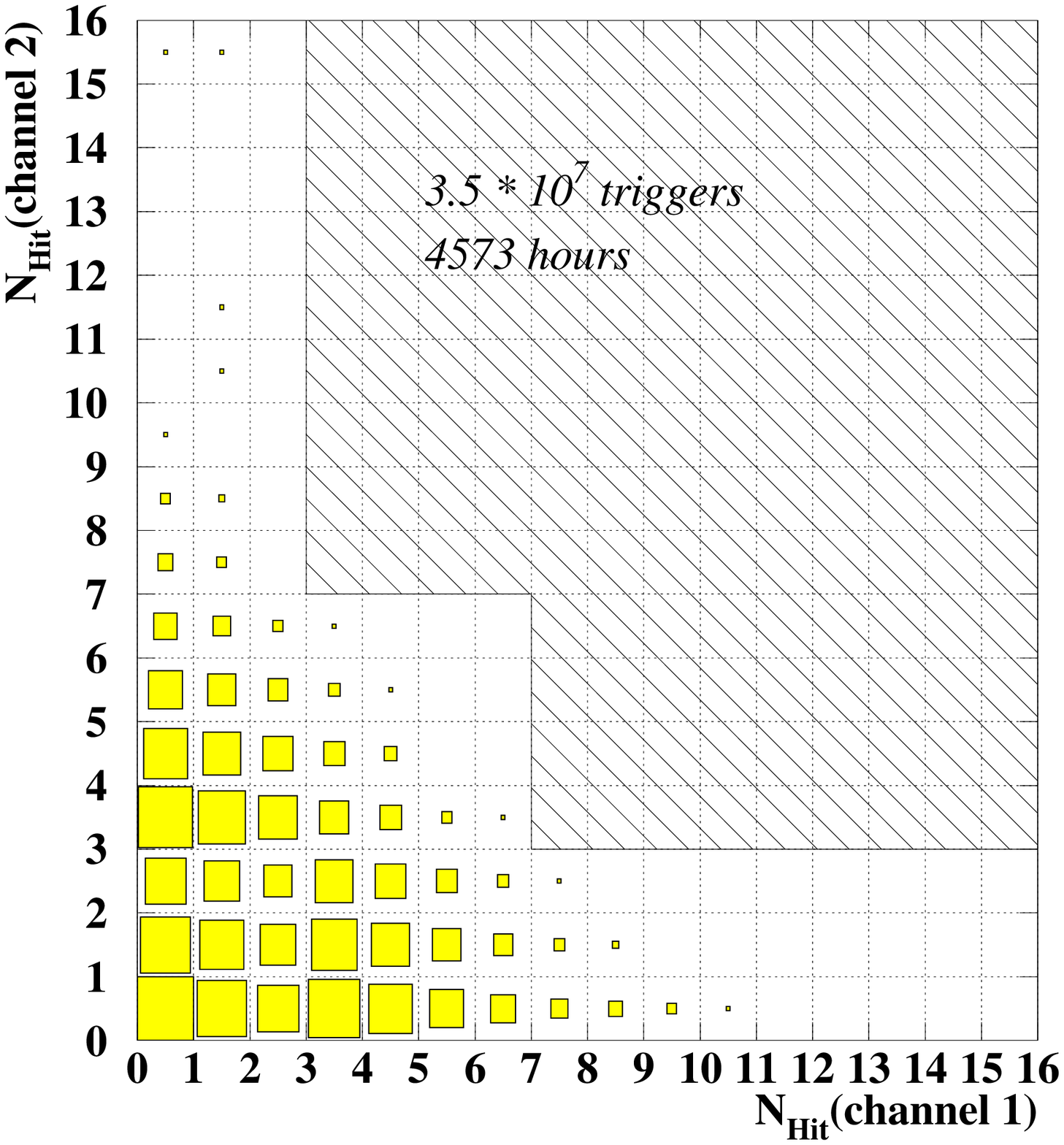,width=7.2cm,height=7.8cm}}
     \parbox[t]{7cm}{
          \caption [1]
         {\footnotesize Hit distributions for a time window of 8 msec,
             as recorded in a 2 hours testrun for channels 14 and 18
             of NT-36. Experimental data are indicated by points,
             the curve gives the Poisson prediction.}
               }
\hspace{.5cm}
     \parbox[t]{7cm}{
          \caption [2]
    {\footnotesize Hit numbers in channel 2 versus
            channel 1 for  the channels of
            all operating $svjaskas$. The shaded region corresponds
            to the off-line trigger condition, see text. }
          }
\end{figure}
\begin{par}
\noindent
of local triggers is described by a Poisson  distribution. If not,
the search for monopoles
via counting rate splashes would become essentially complicated.
It was elucidated that non-Poisson effects are suppressed effectively
by the coincidence between the two PMTs of a pair. This can be seen from
Fig.1. We compare the number of hits detected within 8 msec-time windows
to the calculated distribution. Calculating it, we assumed pure Poisson
fluctuations around the independently measured average hit rate. Such
good agreement with the Poisson assumption is observed for all channels 
with only very few exceptions.
\end{par}

The data taken from April 16th to November 15th 1993 with NT-36 are
analyzed with respect to monopoles  catalyzing
baryon decay. Using the off-line threshold \mbox{$\geq7$ hits}
within \mbox{500 $\mu$sec} we rejected
most events, registrating only the uppermost tail of the
Poisson distribution. To suppress the remaining accidental noise, we
defined an even tighter trigger, requesting that one 
channel had counted \mbox{$\geq7$
hits} and the second channel looking to it's face and situated 
\mbox{7.5 m} away along
string had counted \mbox{$\geq3$ 
hits} in the same \mbox{500 $\mu$sec} (two channels looking face
to face define a {\it svjaska} which is one of the main levels in the
hierarchy of the Baikal array$^1$). This reduces the number of the
experimentally observed monopole candidates to zero. Fig.2 shows the
number of
hits from the \mbox{channel 1} of a svjaska plotted versus the
number of hits  from  the
face-to-face \mbox{channel 2.} 
The product  of  effective  data  taking  time  and
number of \mbox{svjaskas} with both channels operating is 4573 hours for
the investigated period.
\mbox{$3.5\cdot 10^{7}$} monopole triggers with \mbox{$\geq3$ hits} 
have been taken.

\begin{figure}[H]
\mbox{\epsfig{file=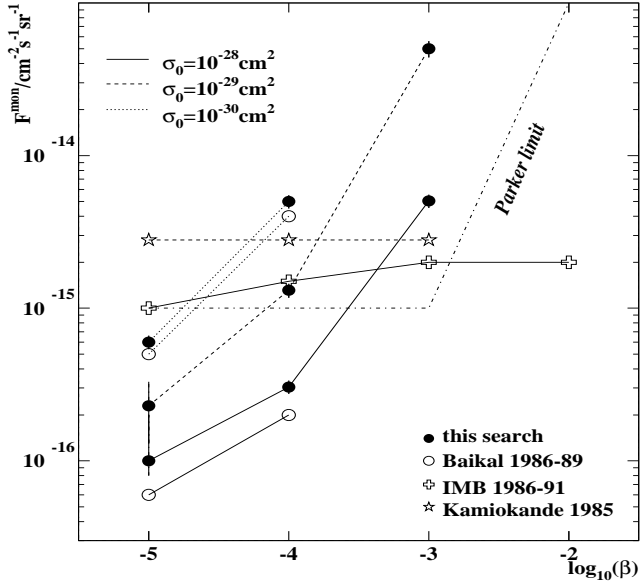,height=7.8cm,width=8.5cm}}
\hspace{0.5cm}
\parbox[b]{6cm}{\caption [3]
{\footnotesize \mbox{Upper limits (90 $\%$ CL)} 
on the natural flux of magnetic
monopoles versus their velocity $\beta$, for different parameters
$\sigma_o$, see text.}}
\end{figure}

After calculating the effective area,
from the non-observation of monopole canditates we obtain the upper
flux limits (90 $\%$ CL)
shown in  Fig.3  together
with our earlier results$^6$, results from IMB$^8$ and KAMIOKANDE$^9$
and with the theoretical Chudakov-Parker
limit$^{10}$. A limit of
$4\cdot 10^{-16}$ cm$^{-2}$ s$^{-1}$ has
been obtained by the Baksan  Telescope$^{11}$ for $\beta > 2 \cdot 10^{-4}$.

The present analysis is rather straight-forward using
the same tight trigger condition for all time periods.
This trigger suppresses fake monopole candidates even
during period with high level of Baikal water
luminescence. It is planned to tune the trigger depending on
the local trigger counting rate. Using this more
sophisticated trigger and the whole statistics
taken for 3 years with NT-36 and NT-72  (652  day's
runtime by December 1995) one can decrease the minimal detectable fluxes
by a factor {$ \approx 10$} compared to the results presented here.
A further considerable progress is expected with
the next stages of NT-200.
\vspace{5mm}
\begin{par}
\noindent
{\bf 3. Search for Fast Monopoles}
\end{par}
\vspace{2mm}

The basic mechanism for light generation by a
relativistic  magnetic monopole
is Cherenkov radiation. The large
magnetic charge results in a giant  light  intensity,
equal to that  of  a  14-PeV  muon  for  relativistic
monopole with  magnetic  charge
\mbox{{$\it g_o = 68.5e$}.}
One  can
search  for  such  monopoles analysing the data
obtained with the muon trigger. Due to the non-stochastical nature of the
Cherenkov light emission of relativistic monopoles (contrary
to a 14 PeV muon!), there
is a close correspondence between
reconstructed monopole track parameters and the number of
hit channels. This can be effectively used  to  select
monopole candidates.
For NT-200, we estimate an effective area roughly as a of
{$\approx 2\cdot 10^{4}$} m$^2$ with respect to
monopoles with $\beta \approx 1$.
More detailed calculations which take into account the background
from fake events produced by atmospheric muons are needed,
but preliminary study
shows that these events can be rejected rather 
effectively.  

The data provided by the muon trigger can be used to
search for monopoles with velocities down to
$\beta \approx 0.1$.
          As was mentioned above in Sect.1, the time and amplitude
information is collected for all  channels
hit within a time window of
           1.8 $\mu$sec around the muon trigger. This time
is comparable with the time needed by particles with
velocities
$\beta \approx 0.1$
to cross the array volume. 
For velocities
$0.6\leq \beta \leq 0.75$,
           most light is due to  Cherenkov  radiation
generated by $\delta$-electrons knocked out by the monopole.
For $\beta \leq 0.6$ the basic mode of light emission
is water luminescence$^{12}$. The  luminescence  of
the Baikal water is extremely small (about one photon
per 5  MeV  energy  loss,  as  we  have studied experimentally
with $\alpha$-particles). Nevertheless,  due
to the giant energy released, the luminescence stimulated
by a monopole with charge
{$\it g_{o}$}
exceeds  the  Cherenkov
light emitted by a relativistic muon down to monopole
velocities of
$\beta \approx 0.1$. The  estimates give
NT-200 effective areas for monopoles with
$\beta = 0.5$ and
$\beta = 0.2$ to be
equal to
                         1000  m$^2$  and
                                           500 m$^2$,
respectively.

We plan to analyse data which taken with NT-36 and NT-72 
as well as data will be obtained with next-stage Baikal arrays 
with respect to 
the search for fast magnetic monopoles.

\vspace{1mm}
\begin{par}
\noindent
{\small
{\it This work was supported at part by the International Science
Foundation (Grants M5M000 and M5M300)
and the Deutscher Akademischer Austauschdienst.}
}
\end{par}

\vspace{2mm}
{\small
{\bf References}
\begin{enumerate}
\vspace{-1mm}
\item I.A.Belolaptikov {\it et  al.}, {\it Proc. 3rd Int. Workshop  on
  Neutrino  Telescopes}, Venice (1991) 365; I.A.Belolaptikov {\it et al.},
  {\it Nucl.Phys.} B {\it (Proc. Suppl.)} {\bf 19} (1991) 375; 
  I.A.Sokalski and Ch.Spiering
  (eds.), {\it The Baikal Neutrino Telescope NT-200}, Baikal note {\bf 92-03};
  I.A.Belolaptikov {\it et al., Nucl. Phys.} B 
  {\it (Proc. Suppl.)} {\bf 43} (1995)  241;
  I.A.Belolaptikov {\it et al.}, {\it Proc. 24th ICRC}, {\bf 1} (1995) 742.
\vspace{-3mm}
\item L.B.Bezrukov {\it et  al.}, paper submitted to these Proceedings.
\vspace{-3mm}
\item V.A.Rubakov, {\it JETP Lett.} {\bf 33} ( 1981) 644.
\vspace{-3mm}
\item C.G.Callan, {\it Phys. Rev.} {\bf D26} (1982) 2058.
\vspace{-3mm}
\item J.Arafune, M.Fukugita, {\it Phys. Rev. Lett.} {\bf 50} (1983) 1901.
\vspace{-3mm}
\item L.B.Bezrukov {\it et al.}, {\it Sov. Journal of Nucl. Phys.} {\bf 52} 
(1990) 54; L.B.Bezrukov, L.A.Kuzmichov, {\it The method of search for slowly
     moving monopole in open water experiments, 2nd
      Winter School of Nonaccelarting Physics}, Zvenigorod, 1983 (unpublished).
\vspace{-3mm}
\item R.Wischnewski {\it et al.}, {\it Proc. 24th ICRC}, {\bf 1} (1995) 658.
\vspace{-3mm}
\item R.Becker-Szendy {\it et al., Phys. Rev.} {\bf D49} (1994) 2162.
\vspace{-3mm}
\item M.Fukugita,  A.Suzuki, {\it Physics
      and Astrophysics of Neutrinos}, (Springer, 1994).
\vspace{-3mm}
\item G.V.Domogatsky, I.M.Zhelesnykh, {\it Sov. Journal of Nucl. Phys.} 
{\bf 10} (1969) 702;
      E.N.Parker, {\it Astrophys. J.} {\bf 160} (1970) 383; M.S.Turner 
{\it et al., Phys. Rev.} {\bf D26} (1982) 1296.
\vspace{-3mm}
\item M.Boliev, private communication.
\vspace{-3mm}
\item I.I.Trofimenko, {\it About one of the possible registration  methods  for
      magnetic monopoles in water Cherenkov detectors}, preprint  INR
    {\bf 765/92} (1992) (in Russian).
\end{enumerate}
}
\end{document}